\documentclass{article}

\usepackage{graphicx}
\usepackage{amsmath,amsfonts,amsthm}
\usepackage{natbib}

\newcommand{\minimize}{\operatornamewithlimits{minimize}}
\newcommand{\maximize}{\operatornamewithlimits{maximize}}

\newcommand{\ev}{\mbox{E}}
\newcommand{\real}{\mathbb{R}}
\title{Ridge Regularization: an Essential Concept in Data Science}
\author{Trevor Hastie\\Statistics Department\\Biomedical Data
  Science Department\\ Stanford University}
\date{January 23, 2020}
\begin{document}
\maketitle
\begin{abstract}
Ridge or more formally $\ell_2$ regularization shows up in many areas
of statistics and machine learning. It is one of those essential
devices that any good data scientist needs to master for their
craft. In this brief {\em ridge fest} I have collected together some
of the magic and beauty of ridge that my colleagues and I have
encountered over the past 40 years
in applied statistics.
\end{abstract}
\section{Ridge and linear regression}
\label{sec:ridge-line-regr}
We first learn of ridge  when we study linear
regression. Ridge provides a remedy for an ill-conditioned $X^\top X$
matrix. If our $n\times p$ regression matrix $X$ has column rank less
than $p$ (or nearly so in terms of its condition number, the ratio of
largest to smallest singular value), then
the usual least-squares regression equation is in trouble:
\begin{equation}
  \label{eq:1}
  \hat\beta=(X^\top X)^{-1}X^\top y.
\end{equation}
The poor (large) condition number  of $X$ is inherited by $X^\top X$, which is either
singular or nearly so, and here we try to invert it. The problem is
that $X^\top X$ has some eigenvalues of zero or nearly zero, so
inverting is not a good idea.
So what we do is add a {\em ridge} on the diagonal --- $X^\top X +\lambda
I$ with $\lambda>0$ ---  which increases all the eigenvalues by $\lambda$
and takes the problem away:
\begin{equation}
  \label{eq:2}
  \hat{\beta}_\lambda= (X^\top X+\lambda I)^{-1}X^\top y.
\end{equation}
This is the ridge regression solution proposed by \cite{HK70} 50 years
ago, and as we will see is alive and strong today.

We can write out the optimization problem that ridge is solving,
\begin{equation}
  \label{eq:3}
  \minimize_\beta\|y-X\beta\|_2^2+\lambda\|\beta\|_2^2,
\end{equation}
where $\|\cdot\|_2$ is the $\ell_2$ (Euclidean) norm. Some simple matrix calculus gets us from \eqref{eq:3} to \eqref{eq:2}.

What value of $\lambda>0$ shall we use? If our main concern is
resolving the numerical issues with solving the least squares
equation, then a small value suffices --- say $\lambda=0.001$ or perhaps
that fraction of the largest eigenvalue of $X^\top X$.

The ridge remedy (\ref{eq:2}) comes with consequences. Under a true
linear model, the ridge estimate is biased toward zero. It also has
smaller variance than the OLS estimate. Selecting $\lambda$ amounts to
a bias-variance trade-off. We address this further in
Section~\ref{sec:ridge-bias-variance}.

The ridge modification  works in many situations where we fit
linear models, and the effect is not as transparent as in
\eqref{eq:2}.
\begin{itemize}
\item With GLMs we model $\eta(x)=\beta^\top x$ and
  $E(y|x)=g(\eta(x))$, and fit by maximum likelihood. If $X$ is
  ill conditioned, this will lead to a Hessian that is flat (near
  zero) in some
  directions, and the Newton algorithm will be unstable.  We can instead maximize
  \begin{equation}
    \ell(\beta;X,y)-\lambda\|\beta\|_2^2,
  \end{equation}
  which, as in (\ref{eq:2}), adds $\lambda I$ to the Hessian, and
  removes the problem.
  One example is logistic regression. Here even if $X$ is well
  behaved, if the classes are separated in $x$-space, the usual
  maximum-likelihood estimator is undefined --- some coefficients are
  infinite. A little bit of ridge resolves the issue.
\item The same is true in the Cox model, multiclass logistic
  regression, and any other model linear in the parameters.
\item In {\em wide-data} situations where $p\gg n$, for example in
  genomics where the variables are SNPs (single nucleotide polymorphisms) and can number in the millions,
  and in document classification in the tens of thousands. Here
  regularization is essential, and $\lambda$ requires careful tuning.  
\end{itemize}
Typically we do not penalize the intercept in the linear model; in the
case of least squares, we can center $X$ and $y$ upfront, and ignore
the intercept. In other GLMs it is handled not quite as simply, but is
a detail which for now we will ignore.

\begin{figure}[hbt]
  \centering
  \includegraphics[trim= 270pt 270pt 54pt
  216pt,clip,width=.3\textwidth]{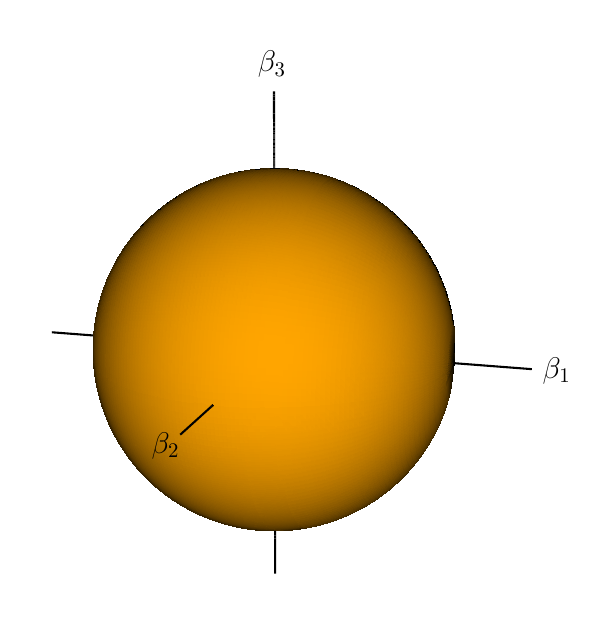}
  \hfil
  \includegraphics[trim= 270pt 270pt 54pt
  216pt,clip,width=.3\textwidth]{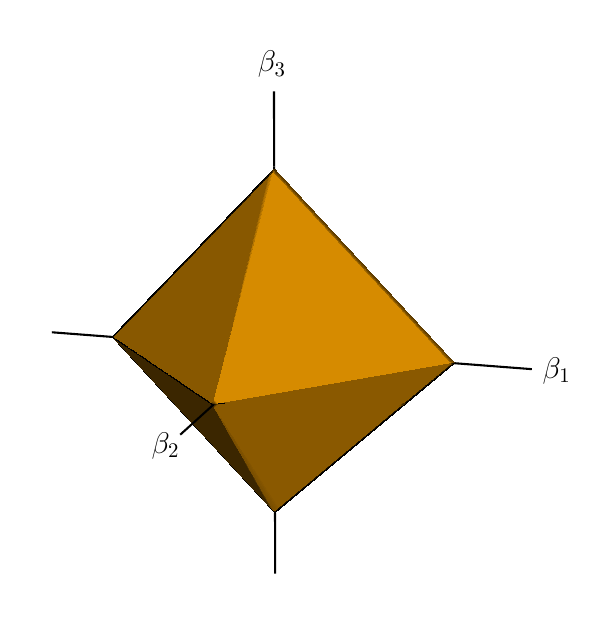}
  \hfil
  \includegraphics[trim= 270pt 270pt 54pt 216pt,clip,width=.3\textwidth]{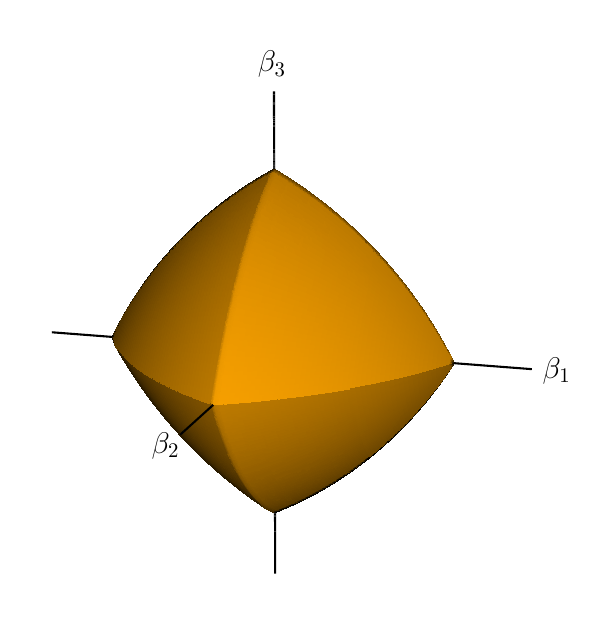}
  \caption{Constraint balls for ridge, lasso and elastic-net
    regularization.
  The sharp edges and corners of the latter two allow for variable
  selection as well as shrinkage.}
  \label{fig:balls}
\end{figure}

We have expressed the ridge problem in {\em Lagrange} form in
\eqref{eq:3}, as opposed to the more evocative {\em bound} form
\begin{equation}
  \label{eq:3a}
  \minimize_\beta\|y-X\beta\|_2 \mbox{ subject to } \|\beta\|_2\leq c.
\end{equation}
The two problems are of course equivalent: every solution
$\hat\beta_\lambda$ to problem \eqref{eq:3} is a solution to
\eqref{eq:3a} with $c=\|\hat\beta_\lambda\|_2$.
The lasso \citep{Ti96} uses an $\ell_1$ rather than an $\ell_2$
penalty in linear models to achieve sparsity:
\begin{equation}
  \label{eq:3las}
  \minimize_\beta\|y-X\beta\|_2^2+\lambda\|\beta\|_1.
\end{equation}
This penalty both shrinks coefficients like ridge, but also sets some
to zero and thus selects. Figure~\ref{fig:balls} compares their
constraint regions, which explains the ability of lasso to set
coefficients to zero. The lasso has inspired ridge-like generalizations,
such as the two mentioned here:
\begin{itemize}
\item {\em elastic net} \citep{zou03:_regres} which mixes the ridge and
  lasso penalty
  \begin{equation}
    \label{eq:26}
    \lambda\left[(1-\alpha)\|\beta\|_2^2+\alpha\|\beta\|_1\right].
  \end{equation}
  It still selects variables like the lasso, but deals more gracefully
  with correlated variables.
\item {\em group lasso} \citep{YL2007} that does lasso-like selection
  for pre-defined groups of variables
  \begin{equation}
    \label{eq:27}
        \lambda\sum_{j=1}^J\|\theta_j\|_2;
      \end{equation}
 here each $\theta_j$ is a vector of parameters. Notice the lack of
 the square on the $\ell_2$ norms, which changes what would be  an overall ridge
 penalty to a group-wise lasso.  The {\em overlap group
   lasso} \citep{jacob09:_group_lasso_with_overl_and_graph_lasso,hastie15:_statis_learn_with_spars}
 allows for hierarchy in selection, and has been used (by us, among
 others) for selecting
 interactions \citep{lim14:_learn_inter_via_hierar_group_lasso_regul}
 and additive models \citep{chouldechova15:_gener_addit_model_selec}.
\end{itemize}

There is a Bayesian view of ridge regression. We assume
$y_i|\beta, X=x_i \sim x_i^\top \beta+\epsilon_i$, with
$\epsilon_i\mbox{ iid } N(0,\sigma_\epsilon^2).$ Here we think of
$\beta$ as random as well, and having a prior distribution
$\beta\sim N(0,\sigma^2_\beta I)$. Then the negative log posterior
distribution is proportional to \eqref{eq:3} with
$\lambda=\sigma_\epsilon^2/\sigma_\beta^2$, and the posterior mean is the
ridge estimator \eqref{eq:2}.  The smaller the prior variance
parameter $\sigma^2_\beta$, the more the posterior mean is shrunk
toward zero, the prior mean for $\beta$.  The Bayesian version of the
lasso uses a Laplace prior, which puts much more mass at and around
zero, and has wider tails than the Gaussian, per unit variance.
Ridge therefore expects more variables in the model, and shrinks them
all to stabilize variance. Lasso expects fewer variables, and hence
is able to shrink those included less to stabilize variance.

\section{Ridge computations and the SVD}
\label{sec:ridge-comp-svd}
In many wide-data and other ridge applications, we need to treat $\lambda$ as a tuning
parameter, and select a good value for the problem at hand. For this
task we have a number of approaches available for selecting $\lambda$
from a series of candidate values:
\begin{itemize}
\item With a validation dataset separate from the training data, we
  can evaluate the prediction performance at each value of $\lambda$.
\item Cross-validation does this efficiently using just the training
  data, and leave-one-out (LOO) CV is especially efficient; see Section~\ref{sec:leave-one-out}.
\item $C_p$ and unbiased risk estimates, where we correct the bias in
  the training risk at each $\lambda$.
\end{itemize}
Whatever the approach, they all
require computing a number of solutions $\hat\beta_\lambda$ at different values of
$\lambda$: the {\em ridge regularization path}.
With squared-error loss as in \eqref{eq:3}, we can achieve great efficiency
via the SVD (singular-value decomposition): $X=UDV^\top$. If $X$ is
$n\times p$, we use the {\em
  full} form of the SVD, with $U$ $n\times n$ orthogonal, $V$ $p\times
p$ orthogonal and $D$ $n\times p$ diagonal, with diagonal entries
$d_1\geq d_2 \geq \cdots\geq d_m\geq 0$, where $m=\min(n,p)$. 
Plugging this into \eqref{eq:2} we get
\begin{equation}
  \label{eq:4}
  \begin{array}{rcl}
    \hat{\beta}_\lambda&=&V(D^\top D+\lambda I)^{-1}D^\top U^\top y\\
    &=&\sum_{d_j>0} v_j \frac{d_j}{d_j^2+\lambda}\langle u_j,y\rangle
  \end{array}
\end{equation}
So once we have the SVD of $X$, we have the ridge solution
for {\emph all} values of $\lambda$.

One can use this representation to show that if two or more variables
are identical (a serious problem for linear regression), their ridge
coefficients are identical (and sum to what would be the ridge
coefficient if just one were included in the model). Similarly,
correlated variables have their coefficients shrunk toward each other
--- the mechanism exploited by the elastic net.

It is also illuminating to look at the fitted values using
\eqref{eq:4}:
\begin{equation}
  \label{eq:5}
  \hat{y}_\lambda=\sum_{d_j>0} u_j \frac{d^2_j}{d_j^2+\lambda}\langle u_j,y\rangle
\end{equation}
The $u_j$ form an orthonormal basis for the column space of $X$. If
$X$ has its column means removed (which we would have done if there
were an unpenalized intercept in the model), then the $u_j $ are the
principal components of $X$.  If $\lambda=0$, then \eqref{eq:5} says
the fit is the projection onto the column space of $X$, using the
principal components as a basis ($\langle u_j,y\rangle$ is the coordinate
of $y$ on the $j$th principal component). With $\lambda>0$ the coordinates of
$\hat y_\lambda$ are shrunk by an increasing amount as we go through
the succession of principal components.

When $n>p$ the ridge solution with $\lambda=0$ is simply the OLS
solution for $\beta$. When $p>n$, there are infinitely many least
squares solutions for $\beta$, all leading to a zero-residual
solution. This is because the least-squares estimating equations
\begin{equation}
  \label{eq:21}
 \nabla_\beta\mbox{RSS}(\beta)= -2X^\top(y-X\beta)=0
\end{equation}
are under determined (more unknowns than equations).
But evidently from \eqref{eq:4} we can get a unique solution
\begin{equation}
  \label{eq:20}
  \hat\beta_{mn}=\sum_{d_j>0} v_j \frac1{d_j}\langle u_j,y\rangle.
\end{equation}
This is the least-squares solution with minimum $\ell_2$ norm.

To see this, we go back to the SVD of $X$ in the $p>n$ setting, and partition
\begin{equation}
  \label{eq:23}
  D_{n\times p}=[\tilde D_{n\times n} : 0_{n\times (p-n)}]
\mbox{ and } V_{p\times p}=
[  \tilde V_{p\times n} :\tilde V^\perp_{p\times (p-n)}].
\end{equation}
Reparametrizing $\beta=V\theta$ and plugging into 
(\ref{eq:21}) we get after some simplification
\begin{equation}
  \label{eq:24}
  U^\top(y-U\tilde D\theta_1)=0,
\end{equation}
where $\theta = [\theta_1:\theta_2]$ is partitioned like $V$.
This defines $\hat\theta_1=\tilde D^{-1}U^\top y$, 
and  $\hat\beta=\tilde V\hat\theta_1 + \tilde V^\perp\theta_2$ for {\em any}
$\theta_2$ is a solution to (\ref{eq:21}). The minimum norm solution
is obtained by setting $\theta_2=0$. 
We have for simplicity assumed that $X$ has full row rank, and hence
$\tilde d_j>0,\;j=1,\ldots,n$; if not we use a generalized inverse $\tilde
D^+$ above, which inserts zeros in the inverse where $\tilde d_j=0$.

\section{Ridge and the bias-variance trade-off}
\label{sec:ridge-bias-variance}
The coefficients of ridge regression are explicitly shrunk toward the
origin. If the data arise from a linear model
\begin{equation}
  \label{eq:29}
  y_i=x_i^\top\beta+\varepsilon_i,\; i=1,\ldots,n
\end{equation}
with i.i.d zero-mean errors $\varepsilon_i$,
then $\hat\beta_\lambda$ will be a biased estimate of $\beta$.
If the $x_i$ are assumed fixed, $n>p$ and $X$ has full column rank, we can get an explicit expression for
this bias from (\ref{eq:4})
\begin{equation}
  \label{eq:31}
    \begin{array}{rcl}
      \mbox{Bias}(\hat\beta_\lambda)&=&E\hat\beta_\lambda-\beta\\
      &=& \sum_{j=1}^p v_j \frac{\lambda}{d_j^2+\lambda}\langle
          v_j,\beta\rangle.
    \end{array}
  \end{equation}
So the coordinates of the true coefficient along different principal
components (of the training data) are differentially shrunk toward
zero --- the smaller the PC, the more the shrinkage. The bias
increases  in all components as  $\lambda$ increases. Note, when
$p>n$, $\hat\beta_\lambda$ lies in the at most $n$-dimensional row
space of $X$, and any component of $\beta$ in the orthogonal component
will contribute to the bias as well.

Similarly there is a nice expression for the covariance matrix under
the sampling model~(\ref{eq:29}):
\begin{equation}
  \label{eq:32}
  \mbox{Var}(\hat\beta_\lambda)=\sigma^2\sum_{j=1}^p\frac{d_j^2}{(d_j^2+\lambda)^2} v_j v_j^\top,
\end{equation}
with $\sigma^2=\mbox{Var}(\varepsilon_i)$.
With $\lambda=0$ this is the usual OLS covariance $\sigma^2(X^\top
X)^{-1}$ in factored form. Here we see that the covariance decreases
uniformly (in a PSD sense) as $\lambda$ increases.

These play off each other when we make predictions at new locations
$x_0$, leading to a bias-variance trade-off.
\begin{equation}
  \label{eq:33}
  \begin{array}{rcl}
    \mbox{MSE}(x_0,\lambda)&=&E(x_0^\top\hat\beta_\lambda-x_0^\top\beta)^2\\
    &=& x_0^\top \mbox{Var}(\hat\beta_\lambda)x_0 + [x_0^\top\mbox{Bias}(\hat\beta_\lambda)]^2
  \end{array}
\end{equation}
For the expected prediction error (EPE) we add $\sigma^2$, the variance of the target response.
\begin{figure}[hbt]
  \centering
  \includegraphics[width=\textwidth]{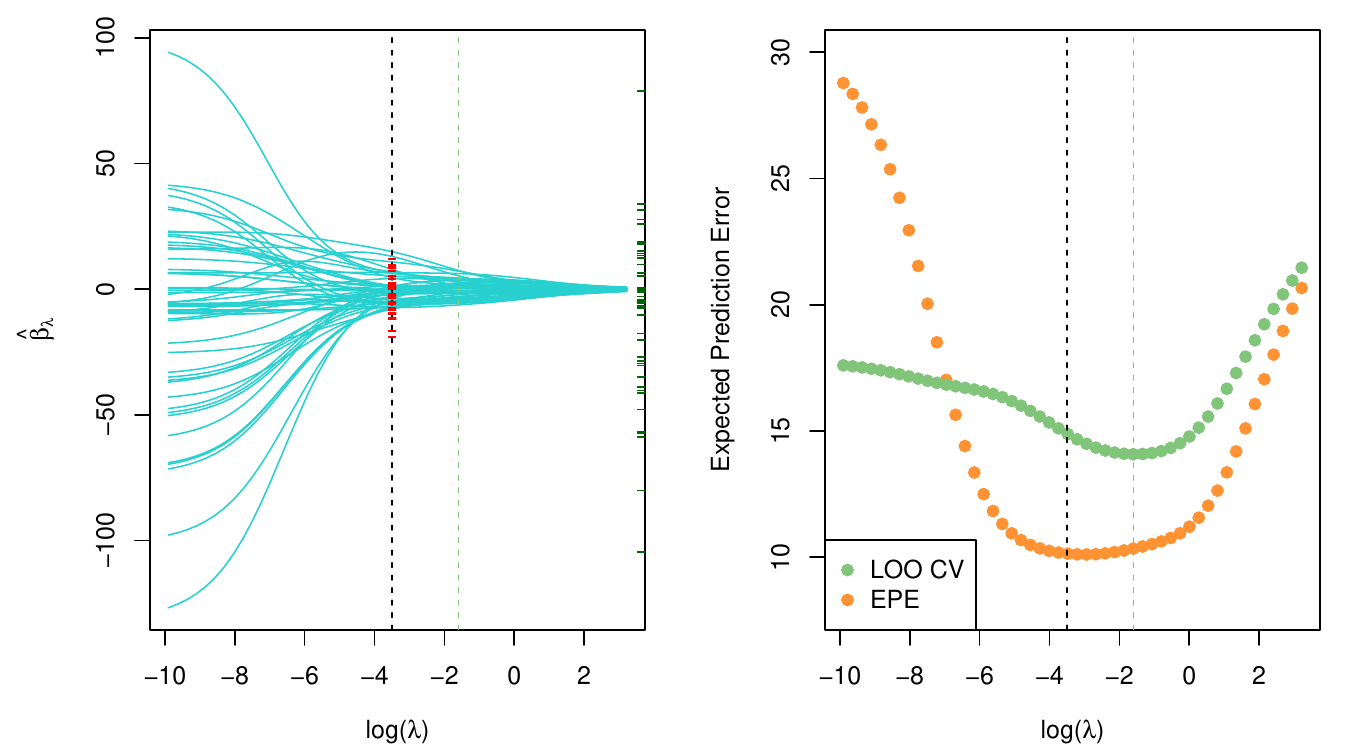}
  \caption{Simulation from a linear model with $n=100$, $p=54$ and
    $SNR=3.3$. [Left panel] Coefficients profiles $\hat\beta_\lambda$
    versus $\log\lambda$. The OLS coefficients are on the far
    left. The left vertical broken line is at the optimal EPE, and the red
    bars are the true coefficients. The second vertical line
    corresponds to the minimum LOO CV error. The dashes on the right axis are
    the James-Stein (uniformly shrunk) estimates. [Right panel] The EPE of the fitted
    model on an infinite test data set (orange):
    $EPE(\lambda)=\sigma^2+E_X(\hat f_\lambda(X)-f(X))^2$, and the
    LOO CV curve estimated from the 100 training points.
    }
  \label{fig:biasvar}
\end{figure}
Even when the model is correct, a shrunken estimate can perform
considerably better than the OLS solution if the bias and variances
are traded off correctly. This is especially true in cases when $p/n$
is large and/or the signal-to-noise ratio is small.
Figure~\ref{fig:biasvar} shows the results of a linear-model simulation with
$n=100$ and $p=54$ and the $SNR=3.3$ (this is a population $R^2$ of
77\%). The OLS coefficients (left
panel, far left in plot) are wild but unbiased. The ridge-shrunken
versions at $\log(\lambda)\approx -4$ are much closer to the true
coefficients (red marks), and predictions using them achieve minimum mean-squared error from
the true linear function $E_X(\hat f_\lambda(X)-f(X))^2$ (right
panel). Of course we don't know the best $\lambda$, and typically use
a left-out data set or cross-validation to estimate the
prediction-error curve empirically. Included in the plot is the
leave-one-out (LOO) CV curve, discussed in
Section~\ref{sec:leave-one-out}, which finds a reasonable value for $\lambda$.

A general form of shrinkage for any multivariate estimator is given by
the celebrated James-Stein formula \citep{james61}. In the context of
linear regression we have
\begin{equation}
  \label{eq:34}
  \hat\beta_{JS}=\left[1-\frac{(p-2)\sigma^2}{\hat\beta^\top X^\top X\hat\beta}\right]\hat\beta,
\end{equation}
where $\hat\beta$ is the OLS estimator, and $\sigma^2$ can be
estimated in the usual fashion via the residual sum-of-squares for
linear regression \citep[Chapter 7]{efron16:_comput_age_statis_infer}.
The left panel of Figure~\ref{fig:biasvar} includes the James-Stein
estimates on the right axis. The corresponding MSE is above 15, and so
is not that good on this example.
\section{Ridge and data augmentation}
\begin{figure}[hbt]
  \centering
  \includegraphics[width=\textwidth]{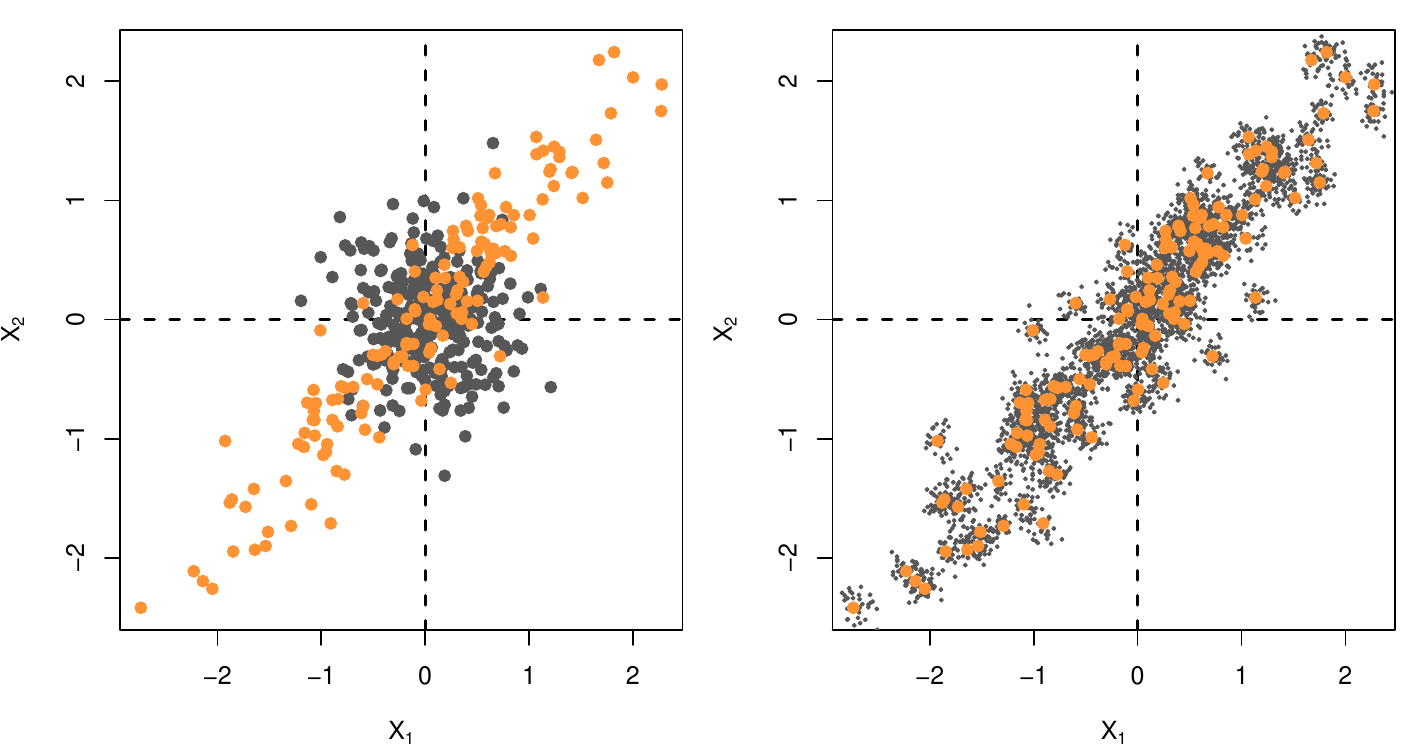}
  \caption{\em Examples of data augmentation. [Left] A mass of fake points at
    the origin fattens the data cloud, and stabilizes the coefficient
    estimates in the direction of the smaller principal axis. [Right]
    Many perturbed versions of the original data points are presented,
    all with the same response. If the perturbations are zero-mean,
    scalar covariance with the right scalar, the result is approximate
  ridge.}
  \label{fig:fatten}
\end{figure}

There are some interesting tricks for fitting a ridge regression using
standard linear-model software.
Suppose we augment $X$ and $y$ in the following way:
\begin{equation}
  \widetilde X = \begin{bmatrix}
    X\\\sqrt{\lambda}I_p
  \end{bmatrix}
  \qquad \tilde y=
  \begin{bmatrix}
    y\\0
  \end{bmatrix}
\end{equation}
Think of the case where $X$ and $y$ are centered. We have added $p$
additional points around the origin, each a distance $\sqrt{\lambda}$
along a coordinate axis. It is straightforward to check that the OLS coefficient is $\hat\beta_\lambda$. Another way to do this
approximately is to augment $X$ in a similar way with $n_a$ random
draws $X_a$ from a $N(0,\lambda I)$ distribution (or any multivariate
distribution with mean zero and covariance $\lambda I$). $y$ is again
padded with $n_a$ zeros. If we fit the model by weighted least
squares, giving weight 1 to the original points, and $1/n_a$ to the
augmentation points, we approximately achieve
$\hat\beta_\lambda$. Approximate because $\frac1{n_a}
X_a^\top X_a
\approx \lambda I$. See the left plot in Figure~\ref{fig:fatten} for
an example.
Another way to achieve this is to perturb each observed data vector
$x_i$ by a random amount. We make $m$ perturbed copies of each
$x_i$: $x'_{ij}=x_i+\varepsilon_{ij},\;j=1,\ldots,m$, where $\varepsilon_{ij} \sim
N(0,\frac\lambda n I)$. Each of the $m$ $x'_{ij}$  vectors gets the same response $y_i$. Then
\begin{equation}
  \label{eq:7}
  \sum_{i=1}^n\sum_{j=1}^{m}x'_{ij}{x'}^\top_{ij}\approx m\left(
    X^\top X + \lambda I\right),
\end{equation}
because the zero-mean cross-terms get averaged away as $m$ gets
large.

Here the augmentation is somewhat artificial. A more realistic
augmentation is used in image classification with deep neural
networks \citep[for example]{10.5555/3217627}. We have a limited number of labeled training images.
The idea is to take each image and apply some natural deformations in
such a way that humans would have no trouble making the
classification. For example, suppose we have an image of a poodle. By
making small rotations, scale and location changes, hue and contrast
changes, it will still look like a poodle. We can now augment our
training set with all these additional versions, creating a cloud of
images around the original, as in Figure~\ref{fig:fatten}[right
panel]. This kind of regularization prevents the model from over-training
on the original image, and by analogy is a form of ridge regularization.

\section{Ridge and dropout regularization}
\label{sec:drop-regul}
Two forms of regularization have arisen in recent years in the
statistical and machine learning communities. They are similar in
spirit to each other, and rather closely aligned with ridge
regression.

Random forests \citep{breiman01:_random_fores} have an option whereby each time a tree is grown, and a
split is contemplated at a node, only a subset $m$ of the $p$
variables available are considered as candidates for splitting.
This causes variables to stand in for each other, or share the
predictive burden ---  a characteristic shared by  ridge regression. The
smaller $m$, the more regularized the fit.

Modern deep neural networks employ a
similar mechanism: {\em dropout learning} \citep{10.5555/2627435.2670313}.
Neural networks have layers of activations or transformations that
feed forward into successive layers through a series of linear models
followed by nonlinear transformations.
For example going from layer $k-1$ with $p_{k-1}$ units to layer $k$
with $p_k$ units,  consider computing the activation $a_\ell^{(k)}$ 
 for a single observation during
  the feed-forward stage of training.
\begin{equation}
  \label{eq:8}
  \begin{array}{rcl}
    z_{\ell}^{(k)}&=&w^{(k-1)}_{\ell
                      0}+\sum_{j=1}^{p_{k-1}}w^{(k-1)}_{\ell j}a^{(k-1)}_j\\
                  a_\ell^{(k)}&=&g(z_\ell^{(k)})
  \end{array}
\end{equation}
 The idea is to randomly set each of the
  $p_{k-1}$ 
activations $a_j^{(k-1)}$
to zero with probability
  $\phi$, and inflate the remaining ones by a factor $1/(1-\phi)$. Hence,
  for this observation, those 
nodes
that survive have to {\em stand
    in} for those omitted. This is done independently for each
  observation, and can also be seen as be a form of ridge
  regularization, and when done correctly improves
  performance. The fraction $\phi$ omitted is
  a tuning parameter, and for deep networks it appears to be
  better to use different values at different layers.

  We illustrate using a simple version of dropout for linear
  regression (see also \cite{NIPS2013_4882}). For simplicity we
  assume all variables have mean zero, so we can ignore
  intercepts. Consider the following random least-squares criterion:
  \begin{equation}
    \label{eq:10}
 L_I(\beta)= \frac12\sum_{i=1}^n\left(y_i-\sum_{j=1}^px_{ij}I_{ij}\beta_j\right)^2.
\end{equation}
Here the $I_{ij}$ are i.i.d  variables  for all $ i,j$ with 
\begin{equation}
  \label{eq:9}
I_{ij}=\left\{
  \begin{array}{rl}
    0&\mbox{with probability $\phi$},\\
    1/(1-\phi)& \mbox{with probability $1-\phi$},
  \end{array}
\right.
\end{equation}
(this particular form is used so that $E[I_{ij}]=1$).
Using simple probability it can be shown that the expected score equations can be written 
\begin{equation}
  \label{eq:escore}
  \ev\left[\frac{\partial L_I(\beta)}{\partial\beta}\right]=-X^\top
  y+X^\top X\beta+\frac\phi{1-\phi} D\beta=0,
\end{equation}
with $D=\mbox{diag}\{\|x_1\|^2,\|x_2\|^2,\ldots,\|x_p\|^2\}$,
where $x_j$ is the $j$th column of $X$.
Hence the solution is given by
\begin{equation}
  \label{eq:deepsol}
  \hat\beta=\left(X^\top X+\frac\phi{1-\phi} D\right)^{-1} X^\top y,
\end{equation}
a generalized ridge regression. If the variables are standardized, the term $D$ becomes a scalar, and the solution is identical to ridge regression.

\section{Ridge and the  kernel trick}
\label{sec:kernel-trick}
We start with the ridge problem \eqref{eq:3} for the case $p\gg n$ and write out the score
equation (up to a factor 2):
\begin{equation}
  \label{eq:13}
  -X^\top(y-X\beta)+\lambda\beta=0.
\end{equation}
For $\lambda>0$ it is easy to see that the solution must satisfy
$\beta=X^\top \alpha$ for some $n$ vector $\alpha$. In other words, the solution
$\hat\beta_\lambda$ lies in the row space of $X$, an $n$-dimensional
subspace of $\real^p$ (we assume $X$ has full row rank, for ease of exposition).

From this we can easily derive
\begin{equation}
  \label{eq:14}
  \hat \alpha_\lambda=(K+\lambda I_n)^{-1}y,
\end{equation}
where $K=XX^\top$ is the $n\times n$ {\em gram} matrix of pairwise
inner products. Likewise the fit vector is given by
\begin{equation}
  \label{eq:15}
  \begin{array}{rcl}
    \hat y_\lambda&=&X\hat\beta_\lambda\\
                  &=& XX^\top \hat \alpha_\lambda\\
     &=&K(K+\lambda I_n)^{-1}y
  \end{array}
\end{equation}
So even though $p$ can be very large, all we need to do are $
n$-dimensional calculations to compute the solution (although the
computations needed to produce $K$ here are $O(pn^2)$). This is 
the {\em kernel trick} in its simplest form.

This is also the case for GLMs, and in fact any linear model fit
obtained by
optimizing a quadratically penalized objective. We will use a GLM as
an example.
Denoting the $n$-vector of fits  by $\eta=X\beta$, we can write the
quadratically penalized log-likelihood problem as
\begin{equation}
  \label{eq:6}
  \maximize_\beta \ell(y,X\beta)-\lambda\|\beta\|_2^2.
\end{equation}
Again it can be shown \citep{hastie03:_pgtn} that the solution has the form $\beta=X^\top \alpha$,
the fit vector $\eta=K\alpha$,
and the optimization in  $\alpha$ becomes
\begin{equation}
  \label{eq:6a}
  \maximize_\beta \ell(y,K\alpha)-\lambda \alpha^\top K \alpha.
\end{equation}

Here the {\em linear} kernel matrix corresponds to ridged linear
models, but the formulation opens the door to the rich world of
function fitting in
reproducing kernel Hilbert spaces (RKHS). For any positive definite
bivariate function ${\cal K}(x,x')$, we can think of it computing
inner products in an implicit feature space $h(x)$: ${\cal
  K}(x,x')=\langle h(x),h(x')\rangle$. We end up fitting functions of
the form $f(x)=\sum_{i=1}^n{\cal K}(x,x_i)\alpha_i$, by solving problems of
the form \eqref{eq:6a} where $K_{ii'}={\cal K}(x_i,x_{i'})$.
See \citet[Chapter 6]{hastie09:_elemen_statis_learn_II} for details.
The support-vector machine for two-class classification is of this
form. With $y_i\in \{-1,+1\}$ and $\eta=K\alpha$, the optimization problem is
\begin{equation}
  \label{eq:16}
  \minimize\sum_{i=1}^n(1-y_i\eta_i)_+ +\lambda \alpha^\top K \alpha.
\end{equation}

For GLMs we can solve \eqref{eq:6a} using standard ridge software.
We compute the Cholesky decomposition of $K=R^\top R$, and
reparametrize via $\theta=R\alpha$ (again an $n$-vector).
Then the objective in $\theta$ becomes
\begin{equation}
  \label{eq:6b}
  \maximize_{\beta'} \ell(y,R^\top \theta)-\lambda\|\theta\|_2^2,
\end{equation}
another ridged linear GLM, but in $n$ rather $p\gg n$ dimensions.
For the linear kernel $K$, it is easy to see that the solution for the
original $\beta$ is $\hat\beta_\lambda=Q\hat{\theta}_\lambda$,
where $X^\top =QR$ is the QR decomposition of $X^\top$ (same $R$).
The beauty here is that we reduce our $n\times p$ wide matrix $X$ to
an $n\times n$ matrix $R^\top$, and then perform our ridge MLE with it
instead. We can perform CV to select $\lambda$ in this space as well,
and need only map back if we want to make predictions on new wide
vectors $x_0$.

\section{Ridge and leave-one-out cross validation}
\label{sec:leave-one-out}
We have already talked of using the SVD to compute the ridge path of
solutions efficiently. This eases the burden when computing the
solution paths $k$ times during $k$-fold cross-validation. For
$n$-fold (LOO) CV, we have another beautiful result for ridge and
other linear operators.
\begin{equation}
  \label{eq:17}
 \mbox{LOO}_\lambda= \sum_{i=1}^n(y_i-x_i^\top\hat\beta_{\lambda}^{(-i)})^2= \sum_{i=1}^n\frac{(y_i-x_i^\top\hat\beta_{\lambda})^2}{(1-R^\lambda_{ii})^2}.
\end{equation}
Here $\hat\beta_{\lambda}^{(-i)}$ is the ridge estimate computed using
the $(n-1)$-observation dataset with the pair $(x_i,y_i)$ omitted, and
\begin{equation}
  \label{eq:18}
  R^\lambda=X(X^\top X+\lambda I)^{-1}X^\top
\end{equation}
is the $n\times n$ ridge operator matrix for the original $n$-observation $X$ matrix. The equation says we can compute all the LOO
residuals for ridge from the original residuals, each scaled up by
$1/(1-R^\lambda_{ii})$.
From \eqref{eq:5} we see we can obtain $R^\lambda$ efficiently for all
$\lambda$ via
\begin{equation}
  \label{eq:19}
  R^\lambda=US(\lambda)U^\top,
\end{equation}
with $S(\lambda)$ the diagonal shrinkage matrix with elements
$d_j^2/(d_j^2+\lambda)$.

One can derive this result using the Sherman-Morrison-Woodbury
identity, but a more general and elegant derivation due to
\cite{GHW79} is as follows.
For each pair $(x_i,y_i)$ left out we are required to solve
\begin{equation}
  \label{eq:11}
  \minimize_\beta\sum_{\ell \neq i}(y_\ell -x_\ell^\top \beta)^2+\lambda\|\beta\|^2
\end{equation}
with solution $\hat\beta_\lambda^{(-i)}$. Let $y_i^*=x_i^\top
\hat\beta_\lambda^{(-i)}$. If we insert the pair $(x_i,y_i^*)$ back
into the size $n-1$ dataset, it will not change the solution to
\eqref{eq:11}, since this point is on the solution surface (and hence
has zero loss at $\beta=\hat\beta_\lambda^{(-i)}$.)
Back at a full $n$ dataset, and using the linearity of the ridge
operator, we  have
\begin{equation}
  \label{eq:12}
  \begin{array}{rcl}
    y_i^*&=&\sum_{\ell\neq i}R^\lambda_{il}y_\ell+R^\lambda_{ii}y_i^*\\
         &=& \sum_{\ell=1}^nR^\lambda_{i\ell}y_i -R^\lambda_{ii}y_i +R^\lambda_{ii}y_i^*,
  \end{array}
\end{equation}
from which we see that $(y_i-y_i^*) =(y_i-\hat y_i)/(1-R^\lambda_{ii})$ with
$\hat y_i=x_i^\top\hat\beta_\lambda$. 

The LOO formula (\ref{eq:17}) appears to break down when $p>n$ in the
limit as $\lambda\downarrow 0$ toward the minimum-norm solution.
It turns out we can get an equally elegant solution in this case as
well \citep{hastie19:_surpr_high_dimen_ridgel_least_squar_inter}.

For $\lambda>0$ from (\ref{eq:15}) we get
\begin{eqnarray}r_\lambda&=&(I_n-K(K+\lambda I_n)^{-1})y\nonumber\\
&=& \lambda (K+\lambda I_n)^{-1}y.\label{eq:six}
\end{eqnarray}
Hence the summands in formula (\ref{eq:17}) can be written as
\begin{equation}\label{eq:three}
y_i-x_i^\top\hat \beta_\lambda^{(-i)} = \frac{\{(K+\lambda
I_n)^{-1}y\}_i}{\{(K+\lambda I_n)^{-1}\}_{ii}}
\end{equation}
(the $\lambda$ multiplier from (\ref{eq:six}) in the numerator and denominator cancel). 
Now we can set $\lambda=0$ to obtain
\begin{equation}
  \label{eq:loomn}
    \mbox{LOO}_{mn}
    =\sum_{i=1}^n\left(y_i-\hat x_i^\top\beta_{mn}^{(-i)}\right)^2
    =\sum_{i=1}^n\left(
      \frac{\{K^{-1}y\}_i}{\{K^{-1}\}_{ii}}
    \right)^2.
\end{equation}

This formula does not apply if there is an unpenalized intercept in
the model. One can show in this case
that the corresponding formula is
\begin{eqnarray}
\mbox{LOO}_{mn} &=& \sum_{i=1}^n\frac{\{\widetilde K^{+}y\}_i}{\{\widetilde K^{+}\}_{ii}},\label{eq:loomnni}
\end{eqnarray}
where $\widetilde K$ is the doubly-centered kernel matrix $\widetilde
K = (I_n-M)K(I_n-M)$ (with rank at most $n-1$), $M=1_n1_n^\top/n$ is the mean projection
operator, and the $\widetilde K^{+}$ is a pseudo inverse.

\section{Ridge, minimum norm and double descent}
\label{sec:minimum-norm-double}
There has been a burst of activity in the machine learning community
over some surprising behavior of a particular class of estimators.
Before we get into detail, a bit of context is in order.
Deep learning models are dominant in certain high SNR domains, such as
image classification. Practitioners have found that fitting a deep
convolutional network with many layers by gradient descent all the
way down to zero training error performs very well on test data. These networks
typically have more parameters than training observations and so this
would normally be classed as severe {\em overfitting}. What has
also been observed is that {\em increasing} the number of parameters
(via more hidden units) can improve performance in some instances
\citep{zhang2016understanding,belkin2018reconciling,hastie19:_surpr_high_dimen_ridgel_least_squar_inter}.

We first make the simple observation \citep{zhang2016understanding}
that with $p\gg n$ and squared-error loss, gradient descent starting
at $\beta=0$ down to zero training error gives the minimum $\ell_2$-norm
solution.  This is because the gradient (\ref{eq:21}) is in the row
space of $X$, and hence all the gradient updates remain there. This
characterizes the minimum-norm solution (see end of Section~\ref{sec:ridge-comp-svd}).
\begin{figure}[hbt]
  \centering
  \includegraphics[width=\textwidth]{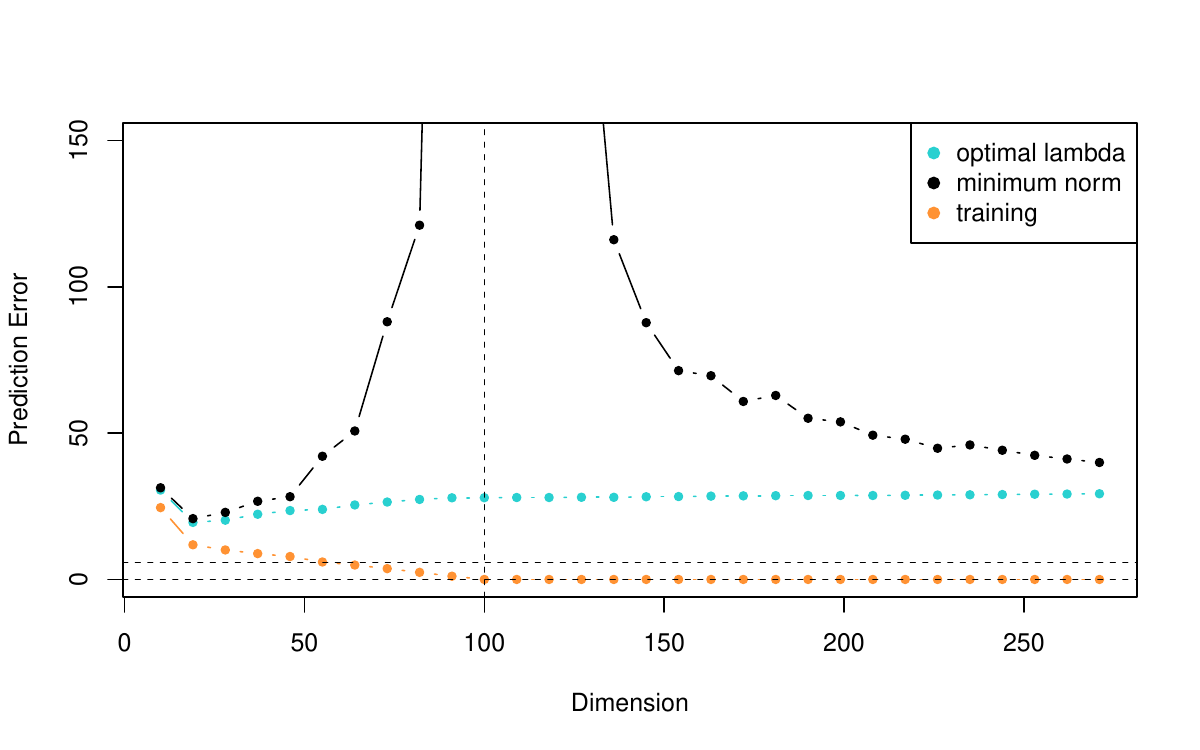}
  \caption{Double descent in the generalization error of minimum-norm
    estimation as the dimension increases.}
  \label{fig:W}
\end{figure}

Figure~\ref{fig:W} is a simple demonstration of this {\em double
  descent} behavior.  We generated data from a
model $y=f(x)+\varepsilon$ with $x\in \real ^9$. The true function $f$
is nonlinear and nonadditive, $x$ is distributed $N(0,I_9)$,
$\varepsilon$ is $N(0,\sigma^2)$, and $\sigma$ is chosen so the SNR
(signal-to-noise ratio) $\mbox{Var} (f(x))/\sigma^2= 3$. The sample
size is $n=100$, and we use a very large test set (10K) to measure
out-of-sample error. For our estimation method we use an additive
model
\begin{equation}
  \label{eq:25}
  g(x;d)=\theta_0+\sum_{j=1}^9\theta_j^\top h_j(x_j;d),
\end{equation}
where each $h_j(\cdot ;d)$ is a $d$-vector of natural-spline basis
functions, with knots chosen at uniform quantiles of the training
values for variable $x_j$ \cite[Chapter
5]{hastie09:_elemen_statis_learn_II}. The total number of parameters
or dimension of the model
is $9d+1$, with $d$ stepping from 1 thru 30. The dimension reaches the
training sample size of $n=100$ when $d=11$.

The black curve shows the test prediction error of the minimum-norm
least squares fits, and the orange curve their training error. The
training error behaves as expected, reaching and staying at zero as
the dimension passes 100. Before 100, the fits are OLS and overfit
almost immediately, and increase to a dramatic level (around 2000)
before descending down again as the dimension increases. The blue
curve corresponds to the optimally tuned ridge estimator which is
fairly flat, with a minimum around dimension 20.

The apparent dilemma here is that the black curve does not show the
usual bias-variance trade-off as the dimension increases. The
explanation for the increase around 100 is that the model has to
interpolate the training data to achieve zero training error. As a
result the prediction curve has to wiggle a lot between training
points, and the $\ell_2$ norm of $\hat \theta$ gets large.
This does not bode well for out-of-sample prediction, since
the test features fall in this in-between region. But as the dimension
grows beyond 100, zero error can be achieved more smoothly with
decreasing norm, and leads to improved out-of-sample
prediction.
So the complexity of the
model is not determined by dimension alone; the potential for smaller
$\ell_2$ norm of the solution at each dimension plays a role as well.
The ``objective'' that is optimized at each dimension combines loss
and complexity (OLS fit with minimum norm), which clouds the usual
bias-variance picture.

\begin{figure}[hbt]
  \centering
  \includegraphics[width=.7\textwidth]{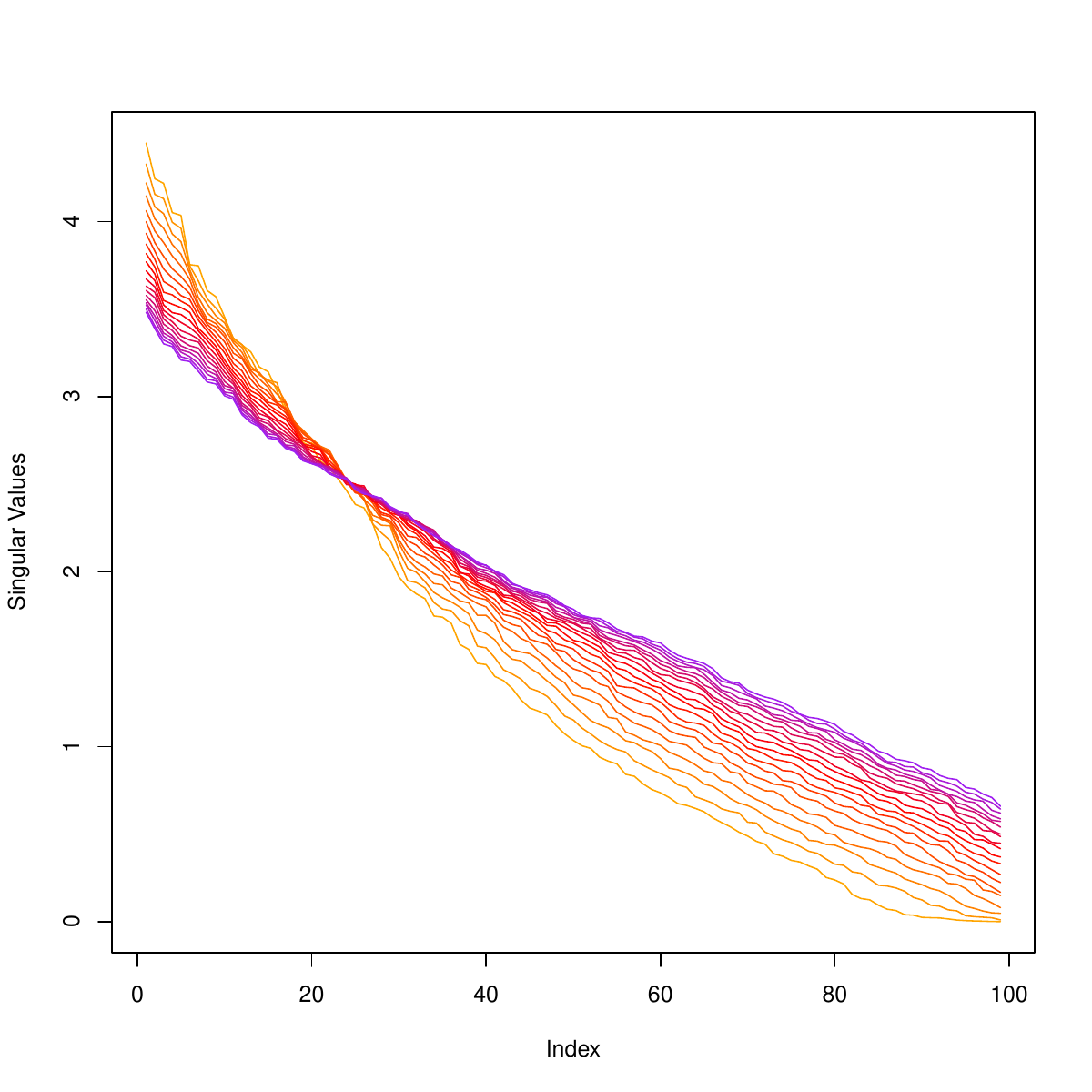}
  \caption{Singular values for the natural spline bases on the
    training data, as the dimension increases from 100 (orange) to 270
  (purple).}
  \label{fig:SV}
\end{figure}

It is interesting to note that the minimum-norm fitting method is
linear in $y$ (see (\ref{eq:20})).  However, as the number of columns (dimension) of the
basis matrix $H$ increases, the small singular values increase leading
to the potential for more stable and smaller $\ell_2$-norm
solutions. Figure~\ref{fig:SV} illustrates this for our example.

\section{Ridge + ridge = lasso}
\label{sec:ridge-+-ridge}
Suppose $X$ is an $m\times n$ matrix.
Consider the optimization problem
\begin{equation}
  \label{eq:28}
  \minimize_{M}\|X-M\|^2_F\quad \mbox{ s.t. rank$(M)\leq q$}.
\end{equation}
Even though the rank constraint makes this problem non-convex, the
solution is well known and easy to compute. We compute the SVD of
$X=UDV^\top$, set all but the top
$q$ singular values of $D$ to zero, and reconstruct.
There is a convex relaxation of this problem \citep{fazel}
\begin{equation}
  \label{eq:28nuc}
  \minimize_{M}\|X-M\|^2_F + \lambda\|M\|_*,
\end{equation}
where $\|M\|_*$ is the {\em nuclear norm} of $M$ --- the sum of the
singular values. The solution is again via the SVD of $X$. Now we
replace the singular values by their soft-thresholded values $\tilde
d_i=(d_i-\lambda)_+$, and reconstruct. This is the lasso version of
rank selection for matrices. Suppose $\lambda$ is such that all but
$q$ of the $\tilde d_i$ are greater than zero, and hence the solution
$\widetilde M_\lambda$ has rank $q$. Now consider the doubly ridged problem \citep{natis-05}
\begin{equation}
  \label{eq:28rr}
  \minimize_{A,B}\|X-AB^\top\|^2_F + \frac\lambda2\|A\|_F^2+\frac\lambda2\|B\|_F^2,
\end{equation}
over matrices $A_{m\times q}$ and $B_{n\times q}$.
This biconvex problem has a solution that coincides with that of
(\ref{eq:28nuc}): $\widetilde A\widetilde B^\top=\widetilde M$. Quite
remarkable that a biconvex $\ell_2$-regularized problem is equivalent to
a convex $\ell_1$ regularized problem!
At this point these connections may seem somewhat academic, since the SVD
of $X$ provides solutions to all the problems above.
These connections show their strength in variations of these problems.
\begin{itemize}
\item If $X$ is massive and sparse, we can compute a low-rank matrix
  approximation by alternating ridge regressions. Given $A$, we obtain
  $B$ via $B^\top=(A^\top A+\frac\lambda2 I_q)A^\top X$, which is a
  dense skinny matrix multiplying a sparse matrix. Likewise for $A$
  given $B$.
\item When $X$ has missing values, we can solve the {\em matrix
    completion} objective
  \begin{equation}
    \label{eq:30}
  \minimize_{M}\|P_\Omega(X-M)\|^2_F + \lambda\|M\|_*,
\end{equation}
where $P_\Omega$ projects onto the observed values of $X$ (i.e. the
Frobenius norm ignores the entries corresponding to missing values in
$X$).
Objective (\ref{eq:30}) is convex in $M$, and
for large $X$ solutions can be obtained again using the alternating
ridged version (\ref{eq:28rr}) \citep{hastie14:_matrix_compl_and_low_rank}.
\end{itemize}

\section{Discussion}
\label{sec:discussion}
We see through these examples that ridge regularization and its
extensions are pervasive in applied statistics.
Although the lasso enjoys widespread popularity in wide-data
scenarios, both ridge and elastic-net claim some of the territory.
In document and web-page classification using bag-of-words and n-grams, synonyms
create problems for methods that select variables. Ridge includes them
all, and suffers less. Furthermore, classifiers are hurt less by bias
than quantitative regressors.
What started off as a simple fix for wayward linear regression models
has evolved into a large collection of tools for data modeling. It would be hard to
imagine the life of a data scientist without them.

\subsubsection*{Acknowledgments}
Thanks to Rob Tibshirani for helpful comments.
This research was partially supported by grants DMS-1407548 and IIS
1837931 from the National Science Foundation, and grant 5R01 EB
001988-21 from the National Institutes of Health.

\bibliographystyle{plainnat}
\bibliography{paper}
\end{document}